\documentstyle[12pt]{article}
\begin{document}
\baselineskip 18pt
\def\today{\ifcase\month\or
 January\or February\or March\or April\or May\or June\or
 July\or August\or September\or October\or November\or December\fi
 \space\number\day, \number\year}
\def\thebibliography#1{\section*{References\markboth
 {References}{References}}\list
 {[\arabic{enumi}]}{\settowidth\labelwidth{[#1]}
 \leftmargin\labelwidth
 \advance\leftmargin\labelsep
 \usecounter{enumi}}
 \def\newblock{\hskip .11em plus .33em minus .07em}
 \sloppy
 \sfcode`\.=1000\relax}
\let\endthebibliography=\endlist
\def\lsim{\ ^<\llap{$_\sim$}\ }
\def\gsim{\ ^>\llap{$_\sim$}\ }
\def\r2{\sqrt 2}
\def\beq{\begin{equation}}
\def\eeq{\end{equation}}
\def\beqn{\begin{eqnarray}}
\def\eeqn{\end{eqnarray}}
\def\rmuu{\gamma^{\mu}}
\def\rmud{\gamma_{\mu}}
\def\PL{{1-\gamma_5\over 2}}
\def\PR{{1+\gamma_5\over 2}}
\def\sinW2{\sin^2\theta_W}
\def\AEM{\alpha_{EM}}
\def\mul{M_{\tilde{u} L}^2}
\def\mur{M_{\tilde{u} R}^2}
\def\mdl{M_{\tilde{d} L}^2}
\def\mdr{M_{\tilde{d} R}^2}
\def\mz2{M_{z}^2}
\def\c2b{\cos 2\beta}
\def\au{A_u}
\def\ad{A_d}
\def\cob{\cot \beta}
\def\v#1{v_#1}
\def\tb{\tan\beta}
\def\epem{$e^+e^-$}
\def\KK{$K^0$-$\bar{K^0}$}
\def\wi{\omega_i}
\def\xj{\chi_j}
\def\Wmu{W_\mu}
\def\Wnu{W_\nu}
\def\m#1{{\tilde m}_#1}
\def\mH{m_H}
\def\mw#1{{\tilde m}_{\omega #1}}
\def\mx#1{{\tilde m}_{\chi^{0}_#1}}
\def\mc#1{{\tilde m}_{\chi^{+}_#1}}
\def\mwi{{\tilde m}_{\omega i}}
\def\mxi{{\tilde m}_{\chi^{0}_i}}
\def\mci{{\tilde m}_{\chi^{+}_i}}
\def\mz{M_z}
\def\sw{\sin\theta_W}
\def\cw{\cos\theta_W}
\def\cb{\cos\beta}
\def\sb{\sin\beta}
\def\rwi{r_{\omega i}}
\def\rxj{r_{\chi j}}
\def\rfp{r_f'}
\def\Kik{K_{ik}}
\def\Fq2{F_{2}(q^2)}
\begin{titlepage}
%\hspace*{10.0cm} OCHA-PP-78

%\hspace*{10.0cm} \today
\  \
\vskip 0.5 true cm 
\begin{center}
{\large {\bf The Chromoelectric and Purely Gluonic Operator Contributions to the 
Neutron Electric Dipole Moment  }}  \\
{\large {\bf  in $N=1$ Supergravity }}
\vskip 0.5 true cm
\vspace{2cm}
\renewcommand{\thefootnote}
{\fnsymbol{footnote}}
 Tarek Ibrahim and Pran Nath
\\
\vskip 0.5 true cm 
\it Department of Physics, Northeastern University  \\
\it Boston, MA 02115, USA  \\
\end{center}

\vskip 4.0 true cm

\centerline{\bf Abstract}
\medskip 
A complete one loop  analysis of the chromoelectric dipole  
contribution to the electric dipole moment (edm) of the quarks and of the 
neutron in N=1 supergravity including the gluino, chargino and neutralino
exchange contributions and exhibiting the dependence  on the two
CP violating phases allowed in the soft SUSY breaking 
sector of minimal supergravity 
models is given. It is found that in significant parts
of the supergravity parameter space the chromoelectric 
dipole  contribution to the neutron edm  is 
comparable to the contribution from the electric dipole term and
can exceed the contribution of the electric dipole term in certain 
regions of the parameter space. 
 An analysis of the contribution
of Weinberg's purely gluonic CP violating dimension six operator within 
supergravity
unification is also given. It is found that this contribution can also be
comparable to the contribution from the electric dipole term in certain 
regions of the supergravity parameter space.   
\end{titlepage}

\newpage 
\section{Introduction}
The electric dipole moment (edm) of fermions is one of the important windows to
new physics beyond the Standard Model (SM). In the SM the edm for the fundamental
fermions arising from the
Kobayashi-Maskawa CP violating phase is much smaller\cite{shaba} than the 
current  experimental limit of $1.1\times 10^{-25}$ecm\cite{exp} 
 and beyond the reach
of experiment in the forseeable future. In supersymmetric unified models new
sources of CP violation arise from the complex phases of the soft SUSY 
breaking parameters which contribute to the edm of the quarks and of the 
 leptons[3-9]. 
 For the case  of the quarks and the neutron there are also the color dipole 
operator
and the CP violating purely gluonic dimension six operator\cite{wein,dai}
which contribute.
However,   
with the exception of the work of ref.\cite{aln} virtually all previous 
analyses of the neutron edm have been done neglecting the contributions of 
these additional operators. 
The reason for this neglect is the presumption\cite{kizu,garisto,falk} that 
their contribution to the neutron edm is small. In the analysis of this 
paper we  show that the contributions of the color dipole operator and 
of the purely gluonic operator are not necessarily small and  
 over a significant 
part of the supergravity parameter space their 
contribution to the neutron edm is comparable 
to  the  contribution from the electric dipole operator and can even 
exceed it in certain regions of the parameter space.

  In this Letter we first derive the full one loop contribution to the color 
  dipole operator arising from the gluino, the chargino and the 
  neutralino exchange 
  contributions exhibiting the dependence on the  two CP violating
  phases allowed in the soft SUSY breaking sector of minimal 
  supergravity models. To our knowledge this is the first complete one loop 
  analysis of this operator. 
  We also recompute the purely gluonic dimension
  six operator to incorporate the dependence on the two CP violating 
  phases. We then give a numerical analysis of 
  the relative strength of the color dipole contribution
  to the neutron edm relative to the contribution from the electric
  dipole term using the framework of supergravity unification under the
  constraint of  radiative breaking of the
  electro-weak symmetry. 
  Our analysis shows that the 
  color dipole contribution relative to the electric dipole contribution 
  can vary greatly over the parameter space of the model. We find 
  that there are significant regions of the parameter space where   
  the color dipole contribution can be comparable to the contribution 
  from  the electric
  dipole term and in some regions of the parameter space the color
  dipole contribution can even exceed the contribution from the electric
  dipole term. A similar analysis is also carried out for the 
  purely gluonic dimension six operator. Here again one finds that there 
  exist regions of the parameter space where the contribution of the
  purely gluonic dimension six operator to the neutron edm 
  can be comparable to, and may even exceed, 
   the contribution from the electric dipole term. Thus one is
   not justified in discarding the effects of the color dipole
   operator and of the gluonic dimension six operator in the neutron 
   edm computation.
   
  The outline of the Letter is as follows: In Sec.2 we give 
  the complete one loop calculation  
  including the gluino, the chargino and the neutralino exchange
  contribution to 
  the color dipole operator and discuss its contribution to the 
  electric dipole moment of the neutron including its dependence on the
  two CP violating phases in minimal supergravity. 
  In Sec.3 we recompute  
   the dimension six gluonic operator to include the effects of the two
   CP violating phases. 
   In Sec.4 we give the renormalization group analysis in
   minimal supergravity of the relative numerical strengths of 
  these contributions and show that they can be comparable to the 
  electric dipole 
  contribution  in significant regions of the parameter space.

\section{Analysis of Chromoelectric Dipole Contribution}
The parameters that enter in minimal supergravity grand unification with
radiative breaking of the electro-weak symmetry can be taken to be 
$m_0, m_{1/2}, A_0, tan\beta$ and phase($\mu$), where 
$m_0$ is the universal scalar  mass, m$_{1/2}$ is the universal gaugino mass,
$A_0$ is the universal trilinear coupling, 
tan$\beta=<H_2>/<H_1>$, where $H_2$ gives mass to the up quark and $H_1$ gives 
mass to the down quark, and $\mu$ is the Higgs mixing 
parameter\cite{cham,applied}. 
As  noted in Sec. 1 only two phases in the soft SUSY breaking sector 
 have a physical meaning in minimal 
supergravity, and we choose them 
to be the  phase  $\alpha_{A0}$ of $A_0$ 
 and the phase $\theta_{\mu0}$ of  $\mu_0$.
The quark chromoelectric dipole moment is defined to be the factor $\tilde d^c$
in the effective operator:
\beq
{\cal L}_I=-\frac{i}{2}\tilde d^c \bar{q} \sigma_{\mu\nu} \gamma_5 T^{a} q
 G^{\mu\nu a}
\eeq
where $T^a$ are the generators of $SU(3)$.
In the following we give an analysis of the one loop contributions 
to  $\tilde d^c$   from 
the chargino, the neutralino and the gluino exchange diagrams
shown in Figs. 1a and 1b.

\subsection{Gluino Contribution}
 The quark-squark-gluino vertex can be derived using the 
 interaction\cite{applied}
\beq
-{\cal L}_{q-\tilde{q}-\tilde{g}}=\r2 g_s T_{jk}^a \sum_{i=u,d}
        (-\bar{q}_{i}^j \PL \tilde{g}_a \tilde{q}_{iR}^k +
        \bar{q}_{i}^j \PR \tilde{g}_a \tilde{q}_{iL}^k) + H.c. ,
\eeq
where $a=1-8$ are the gluino color indices, and $j,k=1-3$ are the quark and
squark color indices. The complex phases enter via the 
squark $(mass)^2$ matrix $ M_{\tilde{q}}^2$ which can be diagonalized by the
transformation
\beq
D_{q}^\dagger M_{\tilde{q}}^2 D_q={\rm diag}(M_{\tilde{q}1}^2,
              M_{\tilde{q}2}^2)                  
\eeq
where  
\beq
\tilde{q}_L=D_{q11} \tilde{q}_1 +D_{q12} \tilde{q}_2
\eeq
\beq
\tilde{q}_R=D_{q21} \tilde{q}_1 +D_{q22} \tilde{q}_2.
\eeq
and where $\tilde{q}_1$ and  $\tilde{q}_2$ are the mass eigenstates.
Writing ${\cal L}$ in terms of $\tilde{q}_1$
 and $\tilde{q}_2$ and  integrating 
out the gluino and squark fields and by using the identities
\beq
T^a_{ij}T^a_{kl}=\frac{1}{2}[\delta_{il}\delta_{kj}-\frac{1}{3}
\delta_{ij}\delta_{kl}]
\eeq
and
\beq
f^{abc} T^b T^c= \frac{3}{2}i T^a
\eeq
one can obtain the gluino exchange contribution to $\tilde d^c$. We find 
\beq
\tilde d_{q-gluino}^c=\frac{g_s\alpha_s}{4\pi} \sum_{k=1}^{2}
     {\rm Im}(\Gamma_{q}^{1k}) \frac{m_{\tilde{g}}}{M_{\tilde{q}_k}^2}
      {\rm C}(\frac{m_{\tilde{g}}^2}{M_{\tilde{q}_k}^2}),
\eeq
where $\Gamma_{q}^{1k}=D_{q2k} D_{q1k}^*$
   and $m_{\tilde{g}}$ is the gluino mass and
\beq
C(r)=\frac{1}{6(r-1)^2}(10r-26+\frac{2rlnr}{1-r}-\frac{18lnr}{1-r}).
\eeq

\subsection{Neutralino Contribution}
Here the CP violating phases enter via the 
 the squark $(mass)^2$ matrix, which contains the phases of  $A_q$ and 
$\mu$, and via the neutralino mass matrix given by  
\beq
M_{\chi^0}=\left(\matrix{\m1 & 0 & -\mz\sw\cb & \mz\sw\sb \cr
                 0  & \m2 & \mz\cw\cb & -\mz\cw\sb \cr
                 -\mz\sw\cb & \mz\cw\cb & 0 & -\mu \cr
                 \mz\sw\sb  & -\mz\cw\sb & -\mu & 0}
                        \right).
\eeq
which carries  the phase of $\mu$. 
The matrix $M_{\chi^0}$ is a complex non hermitian and symmetric matrix,
which can be diagonalized by a unitary transformation such that
\beq
X^T M_{\chi^0} X={\rm diag}(\mx1, \mx2, \mx3, \mx4)
\eeq
Quark-squark-neutralino vertex can be derived from the
interaction\cite{applied}
\beqn
-{\cal L}_{q-\tilde{q}-\tilde{\chi}^{0}} & & = {\sum_{j=1}^{4}
\r2 \bar{u}[(\alpha_{uj}
D_{u11}-\gamma_{uj} D_{u21})\PL}\nonumber\\
& &
{+(\beta_{uj} D_{u11}-\delta_{uj}  D_{u21})\PR]
\tilde{\chi}_j^{0} \tilde{u}_1}
{+\r2 \bar{u}[(\alpha_{uj}
D_{u12}-\gamma_{uj} D_{u22})\PL}\nonumber\\
& &
{+(\beta_{uj} D_{u12}-\delta_{uj}  D_{u22})\PR]
\tilde{\chi}_j^{0} \tilde{u}_2 +(u\rightarrow d)+H.c.}
\eeqn
where $\alpha$, $\beta$, $\gamma$ and $\delta$ are given by
\beq
\alpha_{u(d)j}=\frac{gm_{u(d)}X_{4(3),j}}{2m_W  \sin{\beta}(\cos{\beta})}
\eeq
\beq
\beta_{u(d)j}=eQ_{u(d)}X_{1j}^{'*} +\frac{g}{\cos{\theta_W}}
 X_{2j}^{'*}(T_{3u(d)}-Q_{u(d)}\sin^{2}{\theta_W})
\eeq 
\beq
\gamma_{u(d)j}=eQ_{u(d)}X_{1j}^{'} -\frac{gQ_{u(d)}\sin^{2}{\theta_W}}
{\cos{\theta_W}}  X_{2j}^{'}
\eeq
\beq
\delta_{u(d)j}=\frac{gm_{u(d)}X_{4(3),j}^*}{2m_W  \sin{\beta}(\cos{\beta})}
\eeq
and where
\beq
X_{1j}^{'}=X_{1j} \cos{\theta_W}+X_{2j} \sin{\theta_W}
\eeq
\beq
X_{2j}^{'}=-X_{1j} \sin{\theta_W}+X_{2j} \cos{\theta_W}.
\eeq
The one loop analysis using Eq.(12) gives for $\tilde d^c$
\beq
\tilde d_{q-neutralino}^c=\frac{g_s g^2}{16\pi^2}\sum_{k=1}^{2}\sum_{i=1}^{4}
{\rm Im}(\eta_{qik})
               \frac{\mxi}{M_{\tilde{q}k}^2} 
                {\rm B}(\frac{\mxi^2}{M_{\tilde{q}k}^2}),
\eeq
where
\beqn
\eta_{qik} & &={[-\r2 \{\tan\theta_W (Q_q-T_{3q}) X_{1i}
  +T_{3q} X_{2i}\}D_{q1k}^{*}+
     \kappa_{q} X_{bi} D_{q2k}^{*}]}\nonumber\\
& &
\hspace{4cm} {(\r2 \tan\theta_W Q_q X_{1i} D_{q2k}
     -\kappa_{q} X_{bi} D_{q1k})}.
\eeqn
Here 
\beqn
\kappa_u=\frac{m_u}{\r2 m_W \sb},
 ~~\kappa_{d}=\frac{m_{d}}{\r2 m_W \cb}
\eeqn
where $b=3(4)$ for $T_{3q}=-\frac{1}{2}$($\frac{1}{2})$ and :
\beq
B(r)=\frac{1}{2(r-1)^2}(1+r+\frac{2rlnr}{1-r}).
\eeq
\subsection{Chargino Contribution}
Here the CP violating phases enter via the squark (mass)$^2$ matrix 
and  via the chargino mass matrix  given by
\beq
M_C=\left(\matrix{\m2 & \r2 m_W \sb \cr
        \r2 m_W \cb & \mu }
            \right)
\eeq
which involves the phase of $\mu$. The chargino matrix
can be diagonalized by the unitary transformation:
\beq
U^* M_C V^{-1}={\rm diag}(\mc1, \mc2)
\eeq
The quark-squark-chargino vertex can be derived from the 
interaction\cite{applied} 
\beqn
-{\cal L}_{q-\tilde{q}-\tilde{\chi}^{+}} & & =
{g \bar{u}[(U_{11}
D_{d11}-\kappa_{d}U_{12} D_{d21})\PR}\nonumber\\
& &
{-(\kappa_{u}V_{12}^{*} D_{d11})\PL]
\tilde{\chi}_1^{+} \tilde{d}_1}
{+g \bar{u}[(U_{21}
D_{d11}-\kappa_{d}U_{22} D_{d21})\PR}\nonumber\\
& &
{-\kappa_{u}V_{22}^{*} D_{d11})\PL]
\tilde{\chi}_2^{+} \tilde{d}_1}
{+g \bar{u}[(U_{11}
D_{d12}-\kappa_{d}U_{12} D_{d22})\PR}\nonumber\\
& &
{-(\kappa_{u}V_{12}^{*} D_{d12})\PL]
\tilde{\chi}_1^{+} \tilde{d}_2}
{+g \bar{u}[(U_{21}
D_{d12}-\kappa_{d}U_{22} D_{d22})\PR}\nonumber\\
& &
{-(\kappa_{u}V_{22}^{*} D_{d12})\PL]
\tilde{\chi}_2^{+} \tilde{d}_2}\nonumber\\
& &
{+(u\longleftrightarrow d,
 U\longleftrightarrow V, \tilde{\chi_i}^{+}\rightarrow
\tilde{\chi_i}^{c})+H.c.},
\eeqn
and  for $\tilde d^c$ one gets for the up and down flavors 
\beq
\tilde d_{u-chargino}^c=\frac{-g^2 g_s}{16\pi^2}\sum_{k=1}^{2}\sum_{i=1}^{2}
      {\rm Im}(\Gamma_{uik})
               \frac{\mci}{M_{\tilde{d}k}^2} 
                {\rm B}(\frac{\mci^2}{M_{\tilde{d}k}^2}),
\eeq

\beq
\tilde d_{d-chargino}^c=\frac{-g^2 g_s}{16\pi^2}\sum_{k=1}^{2}\sum_{i=1}^{2}
      {\rm Im}(\Gamma_{dik})
               \frac{\mci}{M_{\tilde{u}k}^2}
                {\rm B}(\frac{\mci^2}{M_{\tilde{u}k}^2}),
\eeq
where 
\beq
\Gamma_{uik}=\kappa_u V_{i2}^* D_{d1k} (U_{i1}^* D_{d1k}^{*}-
                \kappa_d U_{i2}^* D_{d2k}^{*})
\eeq

\beq
\Gamma_{dik}=\kappa_d U_{i2}^* D_{u1k} (V_{i1}^* D_{u1k}^{*}-
                \kappa_u V_{i2}^* D_{u2k}^{*}).
\eeq
The contribution to EDM of the quarks can be computed using the
naive dimensional analysis\cite{manohar} which gives
\beq
d^c_q=\frac{e}{4\pi} \tilde d^c_{q} \eta^c
\eeq
where $\eta^c$ is the renormalization group evolution of the operator 
of Eq.(1) from the electroweak scale down to hadronic scale\cite{brat,aln}.
For the neutron electric dipole moment $d_n$ we use the naive quark model
$d_n=(4d_d-d_u)/3$.
\section{CP Violating Purely Gluonic Dimension 6 operator} 
The gluonic dipole moment $d^G$ is defined to be the factor in the 
effective operator
\beq
{\cal L}_I=-\frac{1}{6}d^G f_{\alpha\beta\gamma}
G_{\alpha\mu\rho}G_{\beta\nu}^{\rho}G_{\gamma\lambda\sigma}
\epsilon^{\mu\nu\lambda\sigma}
\eeq
where $G_{\alpha\mu\nu}$ is the 
 gluon field strength tensor, $f_{\alpha\beta\gamma}$
 are the Gell-Mann coefficients, and $\epsilon^{\mu\nu\lambda\sigma}$
is the totally antisymmetric tensor with $\epsilon^{0123}=+1$.

 Dai et al \cite{dai}
have calculated $d^G$ 
considering the top quark-squark loop with a gluino exchange 
in terms of the complex phase $\phi$ of the gluino mass while the 
squark (mass)$^2$ matrix in their analysis was considered
to be real. In our case the squark (mass)$^2$ matrix is complex and
carries the phases of $A_q$ and of $\mu$. We have recalculated $d^G$ 
for our case and we find 
\beq
d^G=-3\alpha_s m_t (\frac{g_s}{4\pi})^3
{\rm Im} (\Gamma_{t}^{12})\frac{z_1-z_2}{m_{\tilde{g}}^3}
{\rm H}(z_1,z_2,z_t)
\eeq
where
\beq
z_{\alpha}=(\frac{M_{\tilde{t}\alpha}}{m_{\tilde{g}}})^2,
z_t=(\frac{m_t}{m_{\tilde{g}}})^2
\eeq
and  
\beq
 \Gamma_{t}^{12}=D_{t22} D_{t12}^* 
\eeq
where $D_t$ is the diagonalizing matrix for the stop $(mass)^2$ matrix,
and the function H is the same as in \cite{dai}.  
The contribution to $d_n$ from $d^G$ can be estimated by the naive dimensional
analysis\cite{manohar} which gives
\beq
d_{n}^G=\frac{eM}{4\pi} d^G \eta^G
\eeq
where $M$ is the chiral symmetry breaking scale with the numerical value  
1.19 GeV, and $\eta^G$ is the
renormalization group evolution of the operator of Eq.(31) from the electroweak scale
down to the hadronic scale. $\eta^c$ and $\eta^G$ have
 been estimated 
to be $\sim 3.3$\cite{brat,aln}.

\section {RG Analysis and Results}
We  discuss now the relative size of the three contributions to the 
neutron edm, i.e., from the electric dipole operator, from the color
dipole operator and from the purely gluonic   
operator. To study their relative contributions we consider 
a given point in the supergravity parameter space characterized by
the set of six quantities at the GUT scale:
  $m_0,  m_{\frac{1}{2}},  A_0,  \tb,  \theta_{\mu0}$ and
$\alpha_{A0}$. In the numerical
analysis we evolve the gauge coupling constants, the Yukawa couplings, 
magnitudes of the soft SUSY breaking parameters, $\mu$ and the
CP violating phases  from the GUT scale down to the Z boson scale.
We use one-loop renormalization group equations
 (RGEs)  for the soft SUSY breaking parameters
  and two-loop RGEs for the Yukawa and
gauge couplings. Using the data gotten from the RG analysis at  the
scale $M_Z$ we compute the  contributions to the quark edm from the 
electric dipole part ($d^E_q$), from the color dipole part   
($d^C_q$), and from the purely gluonic part ($d^G_n$). 
These are further evolved to the hadronic scale 
by using renormalization group analysis as discussed in Secs. 2 and 3.

\begin{center} \begin{tabular}{|c|c|c|c|c|c|}
\multicolumn{6}{c}{Table~1: $m_{\tilde{g}}
=500$ GeV, $m_0=2000$ GeV, $|A_0|=1.0$} \\
\hline
case &$\tb$ & phases (rad) & $d_n^E(10^{-26} e cm)$  & $d_n^C(10^{-26} e cm)$ 
&$d_n^G(10^{-26} e cm)$\\
\hline
(i) &2 & $\theta_{\mu_0}=0.2$, $\alpha_{A_0}=-0.5$ & 0.124 & 3.049 & -1.168 \\
\hline
(ii) & 2 &  $\theta_{\mu_0}=0.2$, $\alpha_{A_0}=0.5$ & -8.46 & 12.84 & 17.315 \\
\hline
(iii) & 4 & $\theta_{\mu_0}=0.2$, $\alpha_{A_0}=-0.5$ & -4.33 & 0.764 & -5.13 \\
\hline
(iv) & 4 & $\theta_{\mu_0}=0.2$, $\alpha_{A_0}=0.5$ & -11.74 & -12.65 & 7.28 \\
\hline
\end{tabular}
\end{center}

\noindent 

To exhibit the importance of the color dipole operator and 
 of the purely gluonic operator  
we display the relative sizes of $d_n^E, d_n^C$ and of $d_n^G$ for 
few illustrative examples in Table 1.
% Case(i)  of Table 1 shows that $|d_n^C|> |d_n^G| >|d_n^E|$.
  One can understand the
smallness of $d_n^E$  for case(i) in the following way. The main contributions 
to $d_n^E$ arise from the chargino and from the gluino exchange. The chargino
exchange gives a negative contribution while the gluino exchange gives
a positive contribution and the smallness of $d_n^E$ is due to a  
 cancellation between these two. 
 For the color dipole term $d_n^C$ there is also a cancellation 
 but this time the cancellation is only partial and it occurs 
  between the d-quark and the u-quark contributions. 
  Because of a large cancellation for 
 $d_n^E$  and only a partial cancellation for $d_n^C$  
 one has  dominance of $|d_n^C|$ over $|d_n^E|$ in this case.

 One may contrast the result of case(i) with that of case(ii) where
 the sign of $\alpha_{A_0}$ is switched. Here the gluino contribution
 in $d_n^E$ switches sign and this time one has a reinforcement of the 
 chargino and the gluino contributions making $|d_n^E|$ much larger 
 than for case(i). Further, in the color dipole part the d-quark
 contribution in the gluino exchange switches sign and there is  a
 reinforcement of the d-quark and the u-quark contributions making
 $|d_n^C|$ larger than for case(i). We see then that  in this case
  $|d_n^E|$ and $|d_n^C|$ are comparable. One may note that a very 
  large change occurs for the purely gluonic term in going from case(i) to
  case(ii).  To understand the large shift in the value of $d_n^G$
 we display explicitly the imaginary part of Eq. (34)
 
 \beq
{\rm Im}(\Gamma_{t}^{12})=\frac{-m_t}{(M_{\tilde{t}1}^2-M_{\tilde{t}2}^2)}
        (m_0 |A_t| \sin \alpha_{A} + |\mu| \sin \theta_{\mu} \cot\beta),
\eeq
where $\theta_{\mu}$ and  $\alpha_{A}$ are the values of 
$\theta_{\mu_0}$ and of $\alpha_{A_0}$ at the electro-weak scale. From 
Eq. (36) we see that 
the magnitude of ${\rm Im}(\Gamma_{t}^{12})$
 depends on the relative sign and the
magnitude of $\theta_{\mu}$ and of $\alpha_{A}$. Thus one has a cancellation
between the $A_t$ and the $\mu$ terms  when  
 $\theta_{\mu}$ and  $\alpha_{A}$ have opposite signs as in case(i) and 
 a reinforcement between the $A_t$ term  and the $\mu$ term when 
 $\theta_{\mu}$ and $\alpha_{A}$ have the same sign
 as in case(ii). Thus one can qualitatively understand the largeness
 of $d_n^G$ in case(ii) relative to case(i). The very large reduction 
 in $d_n^C$ for case(iii) relative to case (ii) 
 occurs because of a reduction in the down quark
 contribution, which is the dominant term in $d_n$,
  due to a switch in the $\alpha_{A_0}$ sign and a change in 
 the value of tan$\beta$.  In going from case(iii) to case(iv)
 the further switch in the sign of $\alpha_{A_0}$ once again increases the
 down quark contribution making $d_n^C$ the largest in magnitude.
 It is interesting to note that each of the four cases of Table 1 leads 
 to a distinct pattern of hierarchy among the three contributions. 
 Thus one has\\

 \noindent
 (i)  $ |d_n^C|> |d_n^G| >|d_n^E|$,~~
 (ii) $|d_n^G|> |d_n^C| >|d_n^E|$ \\  
 (iii) $|d_n^G|> |d_n^E| >|d_n^C|$,~~
 (iv)$ |d_n^C|> |d_n^E| >|d_n^G|$\\

\noindent
The analysis clearly shows that any one of the three contributions
can dominate $d_n$ depending on the part of the parameter space one is in.
 We note also that $d_n^E$ for all the four cases in Table 1 is consistent 
 with experiment while the total edm  which includes the
 color and the purely gluonic part for cases (ii) and (iv) would be
 outside the experimental bound. 
   
 In Fig. 2  we display the magnitudes of $d_n^E$, $d_n^C$ and $d_n^G$ as a
function of $m_{\tilde{g}}$ for a specific set of $m_0$, $A_0$, $\tb$,
$\alpha_{A_0}$ and $ \theta_{\mu0}$ values. Here we find  that 
$|d_n^C|$  is comparable to $|d_n^E|$ over most of the $m_{\tilde{g}}$ region
and in fact exceeds it for values of $m_{\tilde{g}}$ below $\sim 800$ GeV.
Further $|d_n^G|$ also exceeds $|d_n^E|$ for values of $m_{\tilde{g}}$ below
$\sim 400$ GeV in this case. The broad peak in $d_n^E$ 
arises from a destructive interference  between the gluino
and the chargino components of $d_n^E$ which leads to a relatively rapid
fall off of $d_n^E$ for values of $m_{\tilde g}$ below $\sim 500$
GeV. 
There are 
other regions where similar phenomena occur. To exhibit the commonality
of the largeness of $d_n^C$ we give in Fig.(3a) a scatter plot of the
ratio $|d_n^C/d_n^E|$ for the ranges indicated in the Fig.(3a) caption.
 We see that there exist significant regions of the
parameter space where  the ratio $|d_n^C/d_n^E|\sim O(1)$ 
and hence $d_n^C$ is non-negligible in these regions. 
A scatter plot of $|d_n^G/d_n^E|$ is given in Fig.(3b) and it 
exhibits  a similar phenomenon.

In conclusion we have given the first complete one loop analysis  of the
chromoelectric contribution to the electric dipole moment of the quarks
and of the neutron exhibiting the dependence on the two arbitrary CP 
violating phases allowed in minimal supergravity unification.  
We find that the  relative strength of  the 
chromoelectric contribution varies sharply depending on what part of the
supergravity parameter space one is in. 
In significant parts of the parameter space the chromoelectric contribution
is comparable to the electric dipole contribution and can  even exceed 
it in certain regions of the parameter space. A similar conclusion also 
holds  for the contribution of Weinberg's 
CP violating purely gluonic dimension 
six operator. The analysis of the neutron edm exploring the full parameter
space of supergravity unified models  is outside the scope of
this Letter and will be discussed elsewhere. We only state here
that the inclusion of all the three components of the neutron edm, i.e, $d_n^E,
d_n^C$, and $d_n^G$, is essential in making reliable predictions of the
neutron edm. In fact, the  full analysis shows there exist regions
of the supergravity parameter space where internal cancellation among 
the three components lead to an acceptable value of the neutron
edm without either the use of an excessively heavy SUSY spectrum or an 
excessive 
finetuning of phases. 
These results have important implications for the 
mechanisms needed to suppress the neutron edm in SUSY, for the effect of
CP violating phases on dark matter and on the analyses of baryon 
asymmetry in the universe.

\section {Acknowledgements}
One of us (TI) wishes to thank Uptal Chattopadhyay for discussions and 
help in the RG analysis of this work. This research was supported in part 
by NSF grant number PHY-96020274.

\section{Figure Captions}
Fig. 1a: One loop diagram contributing to the color dipole operator
where the external gluon line ends on an exchanged squark line in
the loop.Squarks are represented by $\tilde q_k$ in  the internal
lines. \\
Fig. 1b: One loop diagram contributing to the color dipole operator
where the external gluon line ends on an exchanged gluino line labelled
by $\tilde g$ in
the loop.  \\	
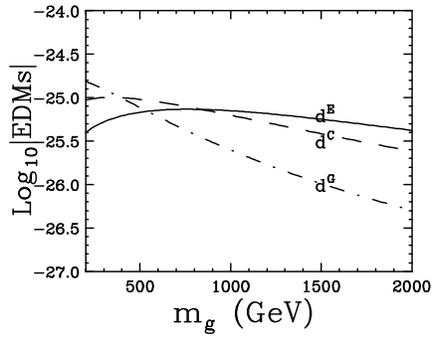
Fig. 2: Plot of $d_n^E$, $d_n^C$, and $d_n^G$ as a function of $m_{\tilde g}$ 
when $m_0$=2000 GeV, tan$\beta$=3.0, $|A_0|$=1.0, $\theta_{\mu_0}$=0.1
and  $\alpha_{A_0}$=0.5.\\
Fig. 3a: Scatter plot of the ratio $|d_n^C/d_n^E|$ as a function of $m_0$
for the case $|A_0|$=1.0, $\tb=2.0$ and the other parameters in the range 
200 GeV$<m_{\tilde g}<600$ GeV
 and $-\pi/5 <\theta_{\mu_0},\alpha_{A_0} <\pi/5$.\\
Fig. 3b: Scatter plot of the ratio $|d_n^G/d_n^E|$ as a function of $m_0$
for the same range of parameters as in Fig. 3a.\\

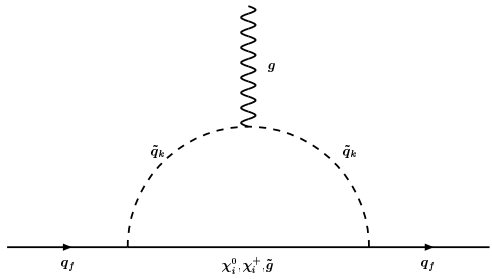
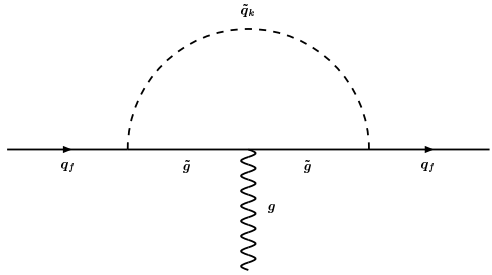
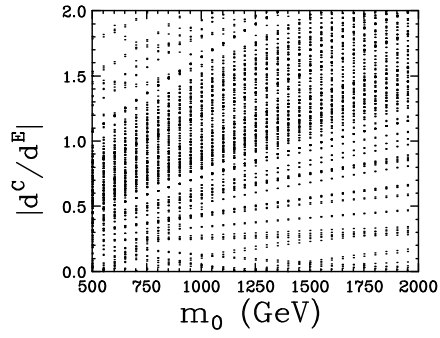
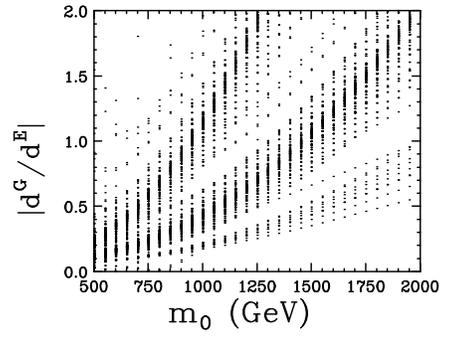
\end{document}